\documentclass [11 pt]{article}

\usepackage{amsmath}
\usepackage{amsfonts}
\usepackage{amssymb}
\usepackage{epsf}
\usepackage{hyperref}

\newcommand\beq{\begin{equation}}
\newcommand\beqn{\begin{equation}\nonumber}
\newcommand\eeq{\end{equation}}
\newcommand\bea{\begin{eqnarray}}
\newcommand\bean{\begin{eqnarray}\nonumber}
\newcommand\eea{\end{eqnarray}}

\def\dR{{\dot R}}
\def\dL{{\dot L}}

\def\bbF{{\mathbb F}}
\def\bbH{{\mathbb H}}

\def\cL{{\mathcal L}}
\def\cR{{\mathcal R}}
\def\cH{{\mathcal H}}
\def\cO{{\mathcal O}}
\def\cF{{\mathcal F}}
\def\cK{{\mathcal K}}

\def\tM{{\widetilde M}}
\def\tR{{\widetilde R}}
\def\tG{{\widetilde G}}
\def\tF{{\widetilde F}}

\def\oP{{\overline P}}
\def\otau{{\overline\tau}}

\begin{document}

\begin{center}{\bf{\Large Classical and Quantum Gravitational Collapse in $d$-dim AdS Spacetime}}
{\bf{\Large  II. Quantum States and Hawking Radiation}}

\bigskip

{Cenalo Vaz$^a$\footnote{e-mail address: Cenalo.Vaz@UC.Edu}, Rakesh Tibrewala$^b$\footnote{e-mail address: rtibs@tifr.res.in} and T.P. Singh$^b$\footnote{e-mail address: tpsingh@tifr.res.in}}

\bigskip

{\it$^a$RWC and Department of Physics, University of Cincinnati,}\\
{\it Cincinnati, Ohio 45221-0011, USA}

\medskip

{\it$^b$Tata Institute of Fundamental Research,}\\
{\it Homi Bhabha Road, Mumbai 400 005, India}

\end{center}

\begin{abstract}
\noindent
In a previous paper we studied the collapse of a spherically symmetric dust distribution (marginally bound LTB) in d-dimensional AdS spacetime and obtained the condition for the formation of trapped surfaces. Here we extend the analysis by giving the canonical theory for the same and subsequently quantize the system by solving the Wheeler-DeWitt equation. We show that for the case of small dust perturbations around a black hole the wave functionals so obtained describe an AdS-Schwarzschild black hole in equilibrium with a thermal bath at Hawking temperature and show the non-trivial dependence of this temperature on the number of spacetime dimensions and the cosmological constant.      
\end{abstract}

\section{Introduction}
One of the fundamental unresolved questions in physics is whether there exists a consistent formulation of quantum theory of gravity and, if it does, whether it cures the singularity problem that plagues the classical theory. Also it is hoped that such a theory, if it exists, will give fundamental insights into the nature of Hawking radiation \cite{hawking} and the origin of black hole entropy as predicted by the use of semiclassical methods \cite{bekenstein1}, \cite{bekenstein2}, \cite{bardeen}. As long as a fully consistent theory of quantum gravity is not available it is worthwhile to study simplified models using different techniques. Among the most useful models are those with spherical symmetry (midisuperspace), since the assumption of spherical symmetry keeps the analysis at a technically simpler level and yet allows one to study various non-trivial issues that are expected to be important in quantum gravity. 

The midisuperspace quantization program has been carried out in a series of papers for spherical dust collapse (LTB collapse) in 3+1 dimensions, both for the marginal \cite{wd}, \cite{vawi1}, \cite{vawi2}, \cite{k1}, \cite{k2} and the non-marginal case \cite{k3}, \cite{k4}, following the work of Kuchar for the Schwarzschild black hole \cite{kuchar}. There one could set up and solve the Wheeler-DeWitt equation for these models and thus obtain exact quantum states, which in the near horizon approximation gave Hawking radiation. However, issues relating to singularity avoidance and to the exact quantum gravitational nature of Hawking radiation could not be addressed fully due to various technical difficulties. To overcome some of the technical difficulties that arise in 3+1 dimensions, the quantum gravitational dust collapse model has also been studied in 2+1 dimensions \cite{sashi1} where it turns out that a black hole solution (the BTZ black hole) is only possible in the presence of a negative cosmological constant. However, in this case the thermodynamics and the statistics of the quantized BTZ black hole are completely different from that of the 3+1 dimensional Schwarzschild black hole. Specifically, whereas the Schwarzschild black hole follows the Boltzmann statistics, the BTZ black hole follows Bose-Einstein statistics \cite{sashi2}. This
is possibly related to the fact that the BTZ black hole has positive specific heat, whereas the specific heat
is negative for the Schwarzschild black hole.

Motivated by this observation we were led to the question: what is the role of the cosmological constant and the number of spatial dimensions in determining the nature of thermodynamics and the statistics of the quantized black hole? The present paper is the second in a series of three papers addressing this question. In the first paper \cite{rtibs} we solved the Einstein equations for a collapsing dust ball in an asymptotically AdS spacetime and examined the nature of the gravitational singularity. Here we look at the canonical formulation of the same which we then quantize obtaining the Wheeler-DeWitt equation, which is then solved on a lattice to obtain exact quantum states. These quantum states are shown to give rise to Hawking radiation in the limit of small dust perturbation around an AdS-Schwarzschild black hole. We also allude to the interplay between the number of spatial dimensions and the value of the cosmological constant and the black hole mass in determining the thermodynamical properties of the black hole. The statistical origin of this dependence based on the present quantum treatment will be taken up in a subsequent paper.   

The paper is organized as follows. In Sec. II we recall the important results concerning the collapse of an inhomogeneous dust cloud under the assumption of spherical symmetry in $d=n+2$ dimensions in the presence of a negative cosmological constant. The canonical formulation for this system is discussed in Sec. III where we also give the contribution(s) to the action coming from the boundary variations along with a discussion of the fall-off conditions. In Sec. IV we make a canonical transformation following Kuchar to obtain a much simpler expression for the Hamiltonian constraint, which is followed by a discussion of the dust action in Sec. V. In Sec. VI we present the quantization of this system which leads to the Wheeler-DeWitt equation. This equation is then solved to obtain exact quantum states on a lattice following DeWitt's regularization scheme. Subsequently we obtain Hawking radiation from these states using a particular ansatz for the inner product. It is remarkable that the $d$-dimensional spherically symmetric system with a negative cosmological constant can be cast in the same canonical form  
as the 3+1 system without a cosmological constant.    

\section{The Classical Solution}

The metric for a spherically symmetric inhomogeneous dust cloud, described by the stress-energy tensor $T_{\mu\nu}=\epsilon(\tau,\rho)U_{\mu}U_{\nu}$, in $d$-dimensional spacetime ($d=n+2$), in the presence of a negative cosmological constant $-\Lambda$ ($\Lambda>0$), in co-moving coordinates is
\begin{equation} \label{ltb metric}
ds^2 = -d\tau^{2}+\frac{\tR^2}{1+2E}d\rho^{2}+R^{2}d\Omega_{n}^{2} \,.
\end{equation}
Here $\Omega_{n}$ is the solid angle for the $n$-sphere, $R=R(\tau,\rho)$ is the radius of the $n$-sphere and $E$, the so-called energy function, is some function of the radial coordinate $\rho$ and we use the notation $\tR=\partial_\rho R$ and $R^*=\partial_\tau R$. For the marginally bound case, corresponding to $E=0$, Einstein equations imply that the energy density $\epsilon$ of the dust cloud and the metric component $R$ evolve as (see \cite{rtibs} for details of the classical solution)
\begin{eqnarray}
\epsilon(\tau,\rho)&=&\frac{(n-1)}{8\pi G}\frac{\tF}{R^{n}\tR}\ , \\
(R^{*})^{2}&=&-\frac{2\Lambda}{n(n+1)}R^{2}+\frac{F(r)}{R^{n-1}}.
\end{eqnarray}
Here $F$, the mass function, is some function of $\rho$ with the condition $F>0$. We note that collapse is described by $R^{*}<0$. Using the scaling freedom in the choice of the radial coordinate $\rho$ to set $R=\rho$ at the initial time $\tau=0$ we can write the solution of the last equation as
\begin{equation} \label{solution for R}
R(\tau,\rho)=\left(\frac{n(n+1)F}{2\Lambda}\right)^{\frac{1}{n+1}}\sin^{\frac{2}{n+1}}\left[\sin^{-1}\sqrt{\frac{2\Lambda\rho^{n+1}}{n(n+1)F}}-\sqrt{\frac{\Lambda(n+1)}{2n}}\tau\right].
\end{equation}
Equation (\ref{solution for R}) shows that for a shell labelled $\rho$ the curvature radius $R$ becomes zero at 
the time
\begin{equation}
\tau_{0}(\rho)=\frac{\sin^{-1}\sqrt{\frac{2\Lambda\rho^{n+1}}{n(n+1)F}}}{\sqrt{\frac{\Lambda(n+1)}{2n}}}.
\end{equation}
This solution can be matched to an exterior Schwarzschild-AdS spacetime 
\begin{equation} \label{exterior metric for negative lambda}
ds^{2}=-\left(1-\frac{F(\rho_{b})}{x^{n-1}}+\frac{2\Lambda x^{2}}{n(n+1)}\right)dT^{2}+\left(1-\frac{F(\rho_{b})}{x^{n-1}}+\frac{2\Lambda x^{2}}{n(n+1)}\right)^{-1}dx^{2}+x^{2}d\Omega^{2}
\end{equation}
where $(T,x)$ are the coordinates in the exterior and $\rho_{b}$ is the boundary of the dust cloud.

\section{Hamiltonian Formalism}
We now want to set up the canonical theory for the general $d=n+2$ dimensional, spherically symmetric collapse problem with and without the cosmological constant. For this we consider the ADM metric
\beq
ds^2 = -N^2 dt^2 + L^2 (dr+N^r dt)^2 + R^2 d\Omega_n^2
\label{adm}
\eeq
where as before $\Omega_n$ is the solid angle 
\beq
d\Omega_n^2 = d\theta_1^2 + \sin^2\theta_1 (d\theta_2^2 + \sin^2\theta_2 (d\theta_3^2 \ldots \sin^2\theta_{n-1}d\theta_n^2)\ldots)).
\eeq
The Einstein-Hilbert action in presence of a negative cosmological constant $-\Lambda$ ($\Lambda>0$), (ignoring the boundary terms), is 
\beq
S_\text{EH} = \frac{n-1}{8n\pi G_d} \int d^dx \sqrt{-g}({}^d\cR+2\Lambda)
\eeq
where ${}^{d}\cR$ is the Ricci Scalar of the $d$-dimensional manifold and $G_{d}$ is the gravitation constant in $d$-dimensions. The above equation after using (\ref{adm}) and integrating over the angular dimensions becomes
\beq
S_\text{EH}=\frac{(n-1)\pi^{\frac{n-1}{2}}}{4n\Gamma(\frac{n+1}{2}) G_d} \int dt \int dr NLR^n({}^{d-1}\cR+ K_{\mu\nu}K^{\mu\nu}-K^2+2\Lambda) = \int dt\int dr~ \cL
\eeq
where ${}^{d-1}\cR$ is the Ricci Scalar of the spatial slice and the prefactor comes from integrating over the solid angle. $K_{\mu\nu}$ is the extrinsic curvature. If we define 
\beq
\tG_d^{-1} = \frac{(n-1)\pi^{\frac{n-1}{2}}}{4n\Gamma(\frac{n+1}2) G_d} ,~~ \Omega_n = \frac{2\pi^{\frac{n+1}2}}{\Gamma\left(\frac{n+1}2\right)}
\eeq 
then
\beq
\cL = \frac 1\tG_d NLR^n({}^{d-1}\cR + K_{\mu\nu}K^{\mu\nu} - K^2+2\Lambda) 
\eeq
or
\bea
\cL &=& -\frac{nR^{n-2}}{\tG_d N} (N^r R'-\dR)[2(N^rLR)'+(n-3)N^r LR'-2R\dL - (n-1)L\dR]\cr\cr
&& -\frac{NR^{n-2}}{\tG_dL^2}[-n(n-1)L^3-2nRL'R'+n(n-1)LR'^2+2nLRR''-2\Lambda R^2L^3]
\eea
From here, we compute the momentum densities,
\bea
P_R &=& \frac{2nR^{n-2}}{\tG_dN}[(N^rLR)'+(n-2)N^rLR'-R\dL-(n-1)L\dR],\cr\cr
P_L &=& \frac{2nR^{n-1}}{\tG_dN} (N^r R'-\dR),
\label{momden}
\eea
which can be solved for the velocities,
\bea
\dR &=& -\frac{\tG_dNP_L}{2nR^{n-1}}+N^rR'\ ,\cr\cr
\dL &=& \tG_d N\left[-\frac{P_R}{2nR^{n-1}}+(n-1)\frac{LP_L}{2nR^n}\right]+(N^rL)'.\ 
\eea
With these definitions the Lagrangian density can be expressed in terms of the canonical variables  
\bea
\cL &=& \tG_d N\left[-\frac{P_LP_R}{2nR^{n-1}}+(n-1)\frac{LP_L^2}{4nR^n}\right]\cr\cr
&& -\frac{NR^{n-2}}{\tG_dL^2}[-n(n-1)L^3-2nRL'R'+n(n-1)LR'^2+2nLRR''-2\Lambda R^2L^3]
\eea
and putting it all together, we compute the Hamiltonian 
\beq
\bbH = \int dr [\dR P_R + \dL P_L - \cL] = \int dr [ N \cH + N^r \cH_r],
\eeq
from which we can read off the Hamiltonain and the diffeomorphism constraints, respectively,
\bea
\cH &=& \tG_d\left[-\frac{P_LP_R}{2nR^{n-1}}+(n-1)\frac{LP_L^2}{4nR^n}\right]\cr
&&+\frac{R^{n-2}}{\tG_dL^2}[-n(n-1)L^3-2nRL'R'+n(n-1)LR'^2+2nLRR''-2\Lambda R^2L^3],\cr\cr
\cH_r &=& R'P_R - LP_L'\ .
\label{constraints}
\eea
It can be shown that this gives the same results for the $n=2$ case as we have derived in our earlier paper.

\subsection{Boundary Variations}

Our next task is to list all possible contributions to the boundary variations. From the diffeomorphism constraint we obtain,
\bea
-\int_{\partial\Sigma}dt N^r P_R \delta R\cr
+\int_{\partial\Sigma}dt N^r L\delta P_L
\eea
and, from the Hamiltonian constraint we get,
\bea
&+&2n \int_{\partial\Sigma}dt \frac{NR^{n-1}}{\tG_d L^2} L'\delta R\cr\cr
&+&2n \int_{\partial\Sigma}dt \frac{NR^{n-1}}{\tG_d L^2} R'\delta L\cr\cr
&-&2n(n-1) \int_{\partial\Sigma}dt \frac{NR^{n-2}}{\tG_d L} R'\delta R\cr\cr
&+&2n \int_{\partial\Sigma}dt \left(\frac{NR^{n-1}}{\tG_d L}\right)'\delta R
\eea
The last boundary variation can be written as the sum of three parts:
\beq
2n \int_{\partial\Sigma}dt \left[\frac{N'R^{n-1}}{\tG_d L}+(n-1)\frac{NR^{n-2}}
{\tG_d L}R'-\frac{NR^{n-1}}{\tG_d L^2}L'\right]\delta R
\eeq
The middle term cancels the third boundary variation of the second list and the 
last term cancels the first boundary variation (in the same list). We are then 
left with just four potential contributions from the boundary, {\it viz.}
\bea
&+&\int_{\partial\Sigma}dt \left[\frac{2nNR^{n-1}}{\tG_d L^2} R'\delta L + 
N^r L \delta P_L + \left(-N^r P_R + \frac{2nN'R^{n-1}}{\tG_d L}\right) \delta R\right]
\label{bdry}
\eea
Whether or not any term contributes depends on the asymptotic conditions we will shortly impose.

\subsection{Equations of Motion}

We now use the equations of motion $\dot{q}=\{q,\bbH\}$, where $q\equiv(R,L,P_{R},P_{L})$, to obtain the evolution equations. Making the choice $N=1$ and $N^{r}=0$ we obtain
\bea \label{rdot}
&& \dot R = \{R,\mathbb{H}\} = \frac{\delta \bbH}{\delta P_R} = -\frac{\tG_d P_L}{2nR^{n-1}}, \\
\label{ldot}
&& \dot L = \{L,\mathbb{H}\} = \frac{\delta \bbH}{\delta P_L} = \tG_d \left[-\frac{P_R}{2nR^{n-1}}+(n-1) \frac{LP_L}{2nR^n}\right], \\
\label{prdot} 
&& \dot P_R = \{P_R, \mathbb{H}\} = -(n-1)\tG_d\left[\frac{P_LP_R}{2nR^n}-
\frac{LP_L^2}{4R^{n+1}}\right] \cr\cr
&&\hskip 0.5cm -\frac{R^{n-3}}{\tG_d L^2}\left[-n(n-1)(n-2)L^3-2n(n-1)RL'R'\right.\cr\cr
&&\left.\hskip 1cm +n(n-1)(n-2)LR'^2+2n(n-1)LRR''+2n\Lambda R^2L^3\right]\cr\cr
&&\hskip 1cm - \frac{2n}{\tG_d}\left(\frac{R^{n-1}L'}{L^2}\right)'+2n(n-1)\left(
\frac{R^{n-2}R'}{\tG_dL}\right)'-2n\left(\frac{R^{n-1}}{\tG_dL}\right)'', \\
\label{pldot}
&& \dot P_L = \{P_L, \mathbb{H}\} = -\frac{\delta \bbH}{\delta L} = -(n-1)\tG_d\frac{P_L^2}{4nR^n}\cr\cr
&&\hskip 0.5cm -\frac{R^{n-2}}{\tG_dL^2}\left[-n(n-1)L^2-n(n-1)R'^2-2n RR''\right.\cr\cr
&&\left.\hskip 2cm -2\Lambda R^2L^2+\frac{4nRL'R'}L\right]-2n \left(\frac{R^{n-2}RR'}{\tG_dL^2}\right)'.
\eea
Taking the time derivative of (\ref{rdot}) and substituting for $\dot{P}_{L}$ we find
\beq \label{ddotr} 
\ddot R = -(n-1)\frac{\dR^2}{2R}-\frac{n-1}{2R}+\frac{(n-1)R'^2}{2RL^2}-\frac{\Lambda R}n.
\eeq
We need to do a similar thing for $L$ and so we consider
\beq
\frac{\dot R'}{R'} = -\frac{\tG_d P_L'}{2nR^{n-1}R'}+(n-1)\frac{\tG_dP_L}{2nR^n}.
\eeq
Using the momentum constraint $\cH_r\approx 0$ to eliminate $P'_{L}$ we obtain
\beq
\frac{\dot R'}{R'} = -\frac{\tG_d P_R}{2nR^{n-1}L}+(n-1)\frac{\tG_dP_L}{2nR^n}
= \frac{\dL}L. 
\eeq
This implies $L=\frac{R'}{f}$ where $f$ is a function only of $r$. If we take $f(r)=\sqrt{1+2E(r)}$ and substitute the expression for $L$ in (\ref{ddotr}) we find
\beq
\ddot R = -(n-1)\frac{\dR^2}{2R}+(n-1)\frac ER-\frac{\Lambda R}n.
\eeq
This equation can be integrated with respect to time 
\beq 
\dR^2 = \frac{F(r)}{R^{n-1}}-\frac{2\Lambda R^2}{n(n+1)}+2E(r).
\label{eqmot}
\eeq
This is seen to be the correct equation for the evolution of $R$ in the non-marginal case. $F(r)$ is of course the mass function and $E(r)$ is the energy function (which is zero for the marginal case).

\subsection{Fall-off conditions}

To determine the fall-off conditions at infinity, we imagine that the metric is smoothly matched to the Schwarzschild-AdS metric at some boundary $r_b$ given by
\beq
ds^2 = -f(r)dt^2+f^{-1}(r)dr^2+r^2d\Omega_n^2
\eeq
where
\beq
f(r)=\left(1-\frac{2\tM}{r^{n-1}}+\frac{2\Lambda r^2}{n(n+1)}\right)
\eeq
where we have set $F(r_b)=2\tM=\tG_d M/n$ and where $M$ is the ADM mass. The fall-off conditions at infinity are given by
($n>1$)
\bea
&& R(t,r) \rightarrow  r + \cO^\infty(r^{-n})\cr\cr
&& L(t,r) \rightarrow \sqrt{\frac{n(n+1)}{2\Lambda}}r^{-1} -\frac 12
\left[\frac{n(n+1)}{2\Lambda}\right]^{3/2}r^{-3} + \cO^\infty(r^{-5})\ldots\cr\cr
&&\hskip 1cm \ldots +\left[\frac{n(n+1)}{2\Lambda}\right]^{3/2} \tM(t)~ r^{-n-2} +\ldots\cr\cr
&& N(t,r) \rightarrow  N_+(t)\left[\sqrt{\frac{2\Lambda}{n(n+1)}}~ r + \cO^\infty(r^{-1})\right]
+\cO(r^{-4})\cr\cr
&& N^r(t,r) \rightarrow \cO^\infty(r^{-n})\cr\cr
&& P_R(t,r) \rightarrow  \cO^\infty(r^{-4})\cr\cr
&& P_L(t,r) \rightarrow \cO^\infty(r^{-2})
\eea
Let's now look at the boundary terms in \eqref{bdry}. Inserting the dependences, we see that the 
contributions amount to
\beq
\frac{2n}{\tG_d}\int_{\partial \Sigma}dt N_+ \left(\delta \tM + \frac{2\Lambda}{n(n+1)}\delta R_1
\right).
\eeq
We will call
\beq
M_+ = \frac{2n}{\tG_d}\left[\tM+\frac{2\Lambda R_1}{n(n+1)}\right]
\eeq
so that the contribution at infinity is 
\beq
\int_{\partial\Sigma}dt N_+\delta M_+
\eeq
and must be {\it subtracted} from the hypersurface action.

The fall off conditions at $r=0$ depend sensitively on the choices we make for the arbitrary functions $E(r)$ and $F(r)$. Near $r=0$, take $E(r) = \sum_{n=0}^\infty E_n r^{n}$ and $F(r)=\sum_{n=0}^\infty F_n r^{n}$, requiring $F_0\geq 0$ (avoiding shell-crossing singularities etc. will involve appropriate conditions on the coefficients). In this way we determine the following fall-off conditions to be met:
\bea
&& R(t,r) \rightarrow a(t) + b(t)r + \cO^0(r^2)\cr\cr
&& L(t,r) \rightarrow \gamma b(t) + \cO^0(r)\cr\cr
&& P_R(t,r) \rightarrow P_{R0}(t) + \cO^0(r)\cr\cr
&& P_L(t,r) \rightarrow P_{L0}(t)+ \cO^0(r)\cr\cr
&& N(t,r) \rightarrow \frac{\gamma N_0(t)}{a^{n-1}} +\cO^0(r^2)\cr\cr
&& N^r(t,r) \rightarrow \cO^0(r)
\eea
where $\gamma=1/\sqrt{1-2E_0}$ and is thus equal to one for the marginally bound case we are considering. They guarantee that the hypersurface action is well behaved at $r=0$. Inserting these into \eqref{bdry} gives
\beq
\frac{2n}{\tG_d}\int_{\partial\Sigma}dt N_0\delta \ln b.
\eeq
If we call $\delta \ln b = -\tG_d\delta M_0/2n$ then we must {\it add} the term
\beq
\int_{\partial\Sigma}dt N_0\delta M_0
\eeq
to the hypersurface action to cancel the contribution from the origin. Combining the two, we must add the term
\beq \label{boundary action}
S_{\partial\Sigma} = -\int_{\partial\Sigma}dt N_+ M_+ + \int_{\partial\Sigma}dt N_0 M_0
\eeq
to the hypersurface action to cancel unwanted boundary variations.

\section{Canonical Transformations}

Our aim now is to write the mass function $F$ in terms of the canonical variables. For this we start by embedding the ADM metric \eqref{adm} in the spacetime described by the LTB metric (\ref{ltb metric}). In what follows we will use the prime and an over-dot to refer to derivatives with respect to the ADM labels $r$ and $t$ respectively and we define $\bar{R}=\tR/\sqrt{1+2E}$. With this a foliation described by $\tau(t,r)$ and $\rho(t,r)$ leads to  
\bea \label{lsquare}
L^2 &=& \bar{R}^2 \rho'^2-\tau'^2, \\
N^r &=& \frac{\bar{R}^2 \dot\rho\rho'-\dot\tau\tau'}{L^2}, \\
N &=& \frac{\bar{R}}L (\dot\tau \rho'-\dot\rho\tau').
\eea
Using these expressions in the momentum densities in \eqref{momden} we find
\beq \label{lpl}
LP_L = \frac{2n}{\tG_d}\frac{R^{n-1}}{\bar{R}(\dot\tau\rho'-\dot\rho\tau')}\left[R'(\bar{R}^2
\dot\rho\rho'-\dot\tau\tau')-\dot R(\bar{R}^2\rho'^2-\tau'^2)\right].
\eeq
Now we note that
\bea
\dot R &=& R^* \dot \tau + \tR \dot\rho = R^* \dot\tau + \bar{R}\sqrt{1+2E}~ \dot\rho,\cr\cr
R' &=& R^* \tau' + \tR \rho' = R^* \tau' + \bar{R} \sqrt{1+2E}~ \rho'.
\eea
Using this along with \eqref{eqmot} in (\ref{lpl}) we find that 
\beq
LP_L = \frac{2nR^{n-1}}{\tG_d\sqrt{1+2E}}\left[\mp R'\sqrt{2E+\frac F{R^{n-1}}-
\frac{2\Lambda R^2}{n(n+1)}} - \left(1-\frac F{R^{n-1}}+\frac{2\Lambda R^2}{n(n+1)}\right)
\tau'\right]
\eeq
where the upper sign is appropriate for expansion and the lower for collapse. Solving for $\tau'$ and calling
\beq
\cF = 1-\frac F{R^{n-1}}+\frac{2\Lambda R^2}{n(n+1)}
\eeq
we have
\beq \label{tau prime in presence of E} 
\tau' = \frac 1\cF \left[\mp R'\sqrt{1+2E-\cF} - \frac{\tG_d\sqrt{1+2E}~LP_L}{2nR^{n-1}}
\right].
\eeq
This is a generalization of the result for $n=2$ in \cite{k3} to arbitrary $n$.

If we now substitute this in (\ref{lsquare}), taking care to maintain 
the appropriate signs for expansion and collapse, we find for either case
\beq
L^2 = \frac{R'^2}{\cF} - \frac{\tG_d^2L^2P_L^2}{4n^2 R^{2(n-1)}\cF}.
\eeq
Solving for $\cF$ we have
\beq
\cF = \frac{R'^2}{L^2} - \frac{\tG_d^2P_L^2}{4n^2 R^{2(n-1)}}
\eeq
and therefore the mass function is obtained in terms of the canonical variables as
\beq
F = R^{n-1}\left[ 1-\frac{R'^2}{L^2} + \frac{\tG_d^2P_L^2}{4n^2R^{2(n-1)}}+\frac{2\Lambda R^2}
{n(n+1)}\right].
\eeq
We would now like to turn the mass function into a canonical variable. Taking the appropriate Poisson brackets shows that 
\beq
P_F=\frac{LP_L}{2R^{n-1}\cF}
\eeq
is canonically conjugate to $F$ ({\it i.e.} $\{F,P_F\}_\text{PB}=1$), which can be seen to be a generalization of the result in earlier work.

We are looking for a coordinate transformation that would take our diffeomorphism 
constraint to 
\beq
\cH_r = R'P_R-LP_L' \rightarrow R'\oP_R + F'P_F
\eeq
{\it i.e.,} we require a new momentum $\oP_R$ such that
\beq
\oP_R = P_R-\frac{LP_L'}{R'} -\frac{F'P_F}{R'} 
\eeq
and the set $\{R,\oP_R,F,P_F\}$ forms a canonical chart on the phase space. We can 
simplify the expression above to get
\beq
\oP_R = P_R - \frac{n-1}2\left[\frac{LP_L}R+\frac{LP_L}{R\cF}\right] - \frac{\Lambda LRP_L}
{n\cF} - \frac{\Delta}{L^2R^{n-1}\cF}
\eeq
where
\beq
\Delta = \frac 1n\left[(R^n)'(LP_L)'-({R^n}')'(LP_L)\right].
\eeq
We may check this expression with the expressions we had in 2+1 and 3+1 dimensions. 
\begin{itemize}
\item For $n=1$ we find
\beq
\oP_R = P_R - \frac{\Lambda LRP_L}{\cF} - \frac{\Delta}{L^2\cF}
\eeq
where 
\beq
\Delta = R'(LP_L)'- R''(LP_L).
\eeq
\item For $n=2$ we have
\beq
\oP_R = P_R - \frac{LP_L}{2R} - \frac{LP_L}{2R\cF} - \frac{\Lambda LRP_L}{2\cF} - 
\frac{\Delta}{L^2\cF}
\label{newmom}
\eeq
where
\beq
\Delta = \frac 1{R}\left[(RR')(LP_L)'-(RR')'(LP_L)\right].
\eeq
\end{itemize}
Both these results were obtained earlier \cite{sashi1}, \cite{k3}.

We now set about showing that the transformation is canonical. Since we already know two coordinates and one conjugate momentum, we simply use the analogues of the standard equations:
\bea
p_i &=& \sum_j P_j {{\partial Q_j} \over {\partial q_i}} + \frac{\partial \bbF}{\partial q_i}\cr\cr
0 &=& \sum_j P_j \frac{\partial Q_j}{\partial p_i} + \frac{\partial \bbF}{\partial p_i}\cr\cr
\cH &=& \cK - \frac{\partial \bbF}{\partial t}
\eea
(where the left hand side are the old phase space variables) {\it i.e.,} the equations
\bea
P_L(r) &=& \int dr' P_F(r') \frac{\partial F(r')}{\partial L(r)} + \frac{\delta
\bbF}{\delta L(r)} \cr\cr
P_R(r) &=& \oP_R(r)+\int dr' P_F(r')\frac{\partial F(r')}{\partial R(r)} + \frac{\delta\bbF}{\delta R(r)}\cr\cr 
0 &=& \int dr' P_F(r') \frac{\partial F(r')}{\partial P_L(r)} + \frac{\delta \bbF}{\delta P_L(r)} \cr\cr
0 &=& \int dr' P_F(r') \frac{\partial F(r')}{\partial P_R(r)} + \frac{\delta \bbF}{\delta P_R(r)}\cr
&&
\label{canonical}
\eea
and ask for a generating function $\bbF[L,R,P_L,P_R]$. The last equation
in (\ref{canonical}) tells us that $\bbF=\bbF[R,L,P_L]$. The third equation gives
\beq
\frac{\tG_d^2P_FP_L}{2n^2R^{n-1}} + \frac{\delta \bbF}{\delta P_L} = 0 \Rightarrow 
\frac{\delta \bbF}{\delta P_L}= -\frac{\tG_d^2LP_L^2}{4n^2 R^{2(n-1)}\left(\frac{R'^2}{L^2}
-\frac{\tG_d^2P_L^2}{4n^2R^{2(n-1)}}\right)}.
\eeq
This gives
\beq
\bbF = \int dr \left[LP_L - \frac{2nR^{n-1}R'}{\tG_d} \tanh^{-1} \frac{\tG_dLP_L}
{2nR^{n-1}R'}\right] + \bbF_1[L,R].
\label{generating}
\eeq
Using this in the first equation \eqref{canonical} then implies
\beq
\frac{\delta \bbF_1}{\delta L}=0
\eeq
showing that $\bbF_1=\bbF_1[R]$. This is a trivial dependence of $\bbF_1$, since we are interested in computing $\oP_R$. We thus take $\bbF_1$ to be independent of $R$, a constant. Calculating $\oP_R$ in (\ref{canonical}) using the form for $\bbF$ in (\ref{generating}) we find that we get the same result as we had before in \eqref{newmom}. Thus we have established that $\{R,F,\oP_R,P_F\}$ form 
a canonical chart.

Finally we want to rewrite the Hamiltonian constraint in \eqref{constraints} in terms of the new variables. Eliminating $L$ and $P_{L}$ in favour of $F$ and $P_{F}$ we get
\bea
\cH &=& -\frac 1{nL}\left[\tG_d \cF P_F \oP_R +\frac{n^2}{\tG_d}\cF^{-1}R'F'\right] \cr\cr
\cH_r &=& R'\oP_R + F'\oP_F
\eea
where it is understood that $L$ is expressed in terms of the new canonical variables. We note that it is remarkable that the constraints have the same form as the constraints in 2+1 dimensions and in 3+1 dimensions. 

\subsection{Boundary Action}

Because varying $N_{+}$ and $N_{0}$ in \eqref{boundary action} would lead to zero ADM mass and restrict $F_{0}$ to zero, respectively, both $N_+$ and $N_0$ should be considered as prescribed functions. By the fall-off conditions, the lapse function, $N^r$, is required to vanish both at the center as well as at infinity. This implies that the time evolution is generated along the world lines of observers with $r=$ constant. If we introduce the proper time of these observers as a new variable, we can express the lapse function in the form $N_+(t) = \dot \otau_+$ and $N_0(t) = \dot \otau_0$. This leads to
\beq
S_{\partial\Sigma} = -\int dt M_+(t)\dot\otau_+ + \int dt M_0(t)\dot \otau_0.
\eeq
Thus we remove the need to fix the lapse function at the boundaries. Extending the treatment in \cite{kuchar}, the aim is to cast the homogeneous part of the action into Liouville form, finding a transformation to new variables that absorb the boundary terms. This can be done by introducing the mass density $\Gamma =nF'/\tG_{d}$ as a new canonical variable. Define
\beq \label{definition of F}
F(r) = \frac{\tG_d}{n}M_0 + \frac{\tG_d}{n}\int_0^r dr'\Gamma(r),
\eeq
and reconsider the Liouville form
\bea
\Theta &:=& - M_+ \delta \otau_+ + M_0 \delta\otau_0+\int_0^\infty dr P_F \delta F\cr\cr
&=&  \otau_+ \delta M_+ - \otau_0 \delta M_0+\int_0^\infty dr P_F \delta F,
\eea
where we have dropped an exact form. But
\beq
\delta F = \frac{\tG_d}{n}\left[\delta M_0 +\int_0^r dr' \delta \Gamma (r')\right]
\label{Fdef}
\eeq
gives
\beq
\Theta = \left(\frac{\tG_d}{n}\int_0^\infty dr' P_F(r') - \otau_0\right)\delta M_0 + \frac{\tG_d}{n}\int_0^\infty dr P_F(r) \int_0^r dr' \delta \Gamma(r') + \otau_+ \delta M_+.
\eeq
Noting further that\footnote{See Kucha\v r \cite{kuchar}. Consider
\bean
&&\left(\int_0^r dr' \delta \Gamma(r') \times \int_{r}^\infty dr' P_F(r')\right)'\cr\cr
= && \delta \Gamma(r) \times \int_r^\infty dr' P_F(r')-P_F(r)\times\int_0^r dr'
\delta \Gamma(r')
\eea
Integrating the left hand side from $0$ to $\infty$ gives zero, therefore
\beqn
\int _{0}^\infty dr  P_F(r) \int_0^r dr' \delta \Gamma(r')=\int_{0}^\infty
dr~ \delta\Gamma(r)\int_r^\infty dr' P_F(r')
\eeq
}
\beq
\int _{0}^\infty dr  P_F(r) \int_0^r dr' \delta \Gamma(r')=\int_{0}^\infty
dr~ \delta\Gamma(r)\int_r^\infty dr' P_F(r'),
\eeq
we can write the Liouville form as
\bea
\Theta &=& \left(\frac{\tG_d}{n}\int_0^\infty dr' P_F(r') - \otau_0\right)\delta M_0\cr\cr
&&\hskip 1cm  + \frac{\tG_d}{n}\int_0^\infty dr \delta \Gamma(r) \left(\int_0^\infty dr P_F(r) -
\int_0^r dr' P_F(r')\right)+ \otau_+ \delta M_+ \cr\cr
&=& \left(\frac{\tG_d}{n}\int_0^\infty dr' P_F(r') - \otau_0\right)\delta M_0\cr\cr
&&\hskip 1cm  + \frac{\tG_d}{n}(\delta M_+-\delta M_0)\int_0^\infty dr P_F(r) - \frac{\tG_d}{n}
\int_0^\infty dr 
\delta \Gamma(r) \int_0^r dr' P_F(r')+ \otau_+ \delta M_+ \cr\cr
&=&p_0\delta M_0 + p_+ \delta M_+ + \int_0^\infty dr P_\Gamma(r) \delta \Gamma(r),
\eea
where
\bea
&& p_0 = -\otau_0,\cr\cr
&&p_+ = \otau_+ + \frac{\tG_d}{n}\int_0^\infty dr P_F(r),\cr\cr
&&P_\Gamma(r) = - \frac{\tG_d}{n}\int_0^r dr' P_F(r').
\label{congamma}
\eea
The new form of the action is then
\beq
S_\text{EH} = \int dt \left(p_0 \dot M_0 + p_+ \dot M_+ + \int dr~ [\oP_R\dot R + P_\Gamma
\dot\Gamma - N \cH^g - N^r \cH_r] \right),
\eeq
where the new constraints read
\bea
&&\cH^g = -\frac 1{L}\left[-\cF \oP_R P_\Gamma'+ \cF^{-1} R'\Gamma \right]\approx 0,\cr\cr
&&\cH_r = R' \oP_R -\Gamma P_\Gamma' \approx 0.
\eea
Notice that $\tG_d$ (and $n$) dependence has been absorbed in the definitions of $\Gamma$ and $P_\Gamma$. In this way we arrive at the remarkable result that the constraints have the same form in any dimension. So far we have not included the dust action. We do this below.

\section{Dust Action}

In this section, we rewrite the Hamiltonian constraint in quadratic form including the dust action. For this we begin with the dust action (see \cite{torre}, \cite{brown} for details on the dust action)
\beq
S^d = -\frac 12 \int d^dx \sqrt{-g}~ \varepsilon(x)[g_{\alpha\beta} U^\alpha U^\beta+1] = 
-\frac{\Omega_n}2 \int dt\int dr NLR^n \varepsilon(x) [g^{\alpha\beta} U_\alpha U_\beta +1].
\eeq
We only consider non-rotating dust and use
\beq
U_\mu = -\tau_{,\mu}
\eeq
where $\tau$ is the dust proper time. Using the ADM metric for spherical symmetry we find
\beq
S^d = -\frac{\Omega_n}2 \int dt\int dr NLR^n\varepsilon(t,r)\left[-\frac{{\dot\tau}^2}{N^2}+
\frac{2N^r}{N^2}\dot\tau\tau'+\left(\frac 1{L^2}-\frac{{N^r}^2}{N^2}\right)\tau'^2+1
\right]
\eeq
and it follows that
\beq
P_\tau=\frac{\delta \cL}{\delta\dot\tau}=\frac{\Omega_n NLR^n\varepsilon}{N^2}
[\dot\tau-N^r\tau'].
\eeq
Thus
\beq
\dot\tau=\frac{NP_\tau}{\Omega_n LR^n\varepsilon}+N^r\tau'
\eeq
and the Hamiltonian can be written as
\beq
\cH = N\cH^d + N^r \cH_r^d
\eeq
with
\bea
&& \cH^d = \frac{P_\tau}L\sqrt{\tau'^2+L^2},\cr\cr
&& \cH^d_r = \tau'P_\tau.
\eea
Thus we have the full gravitation+dust constraints
\bea
&&\cH^g = -\frac 1{L}\left[-\cF \oP_R P_\Gamma'+ \cF^{-1} R'\Gamma - P_\tau
\sqrt{\tau'^2+L^2}\right]\approx 0, \\
\label{final diffeo constraint}
&&\cH_r = \tau'P_\tau + R' \oP_R -\Gamma P_\Gamma' \approx 0.
\eea
We can use the diffeomorphism constraint to simplify the Hamiltonian constraint. Doing so we finally obtain the following expression for the Hamiltonian constraint which, because of the absence of any derivative terms, is much easier to use for quantization 
\beq \label{final hamiltonian constraint}
\cH^{g}=P_\tau^2 + \cF\oP_R^2 -\frac{\Gamma^2}{\cF}\approx 0.
\eeq
This is precisely the same form that we had in the 3+1 case; thus quantization 
can proceed as before.

\section{Quantum States and Hawking Radiation}

We now quantize the diffeomorphism constraint \eqref{final diffeo constraint} and the Hamiltonian constraint \eqref{final hamiltonian constraint}. For this we make the substitutions $P_{\tau}\rightarrow -i\hbar\frac{\delta}{\delta\tau(r)},\  \bar{P}_{R}\rightarrow -i\hbar\frac{\delta}{\delta R(r)},\  P_{\Gamma}\rightarrow -i\hbar\frac{\delta}{\delta\Gamma(r)}$.
With this the quantized form of the diffeomorphism constraint is
\begin{equation} \label{quantized diffeo}
-i\hbar\left[R'\frac{\delta}{\delta R(r)}+\tau'\frac{\delta}{\delta\tau(r)}-\Gamma\left(\frac{\delta}{\delta\Gamma(r)}\right)'\right]\Psi[\tau(r'),R(r'),\Gamma(r')]=0.
\end{equation}
The quantized version of the Hamiltonian constraint, the Wheeler-DeWitt equation for the system, is 
\begin{equation} \label{wdw equation}
\left[\left(-i\hbar\frac{\delta}{\delta\tau(r)}\right)^{2}+\mathcal{F}\left(-i\hbar\frac{\delta}{\delta R(r)}\right)^{2}-\hbar^{2}A(R,F)\delta(0)\frac{\delta}{\delta R(r)}-\hbar^{2}B(R,F)\delta(0)^{2}-\frac{\Gamma^{2}}{\mathcal{F}}\right]\Psi[\tau(r'),R(r'),\Gamma(r')]=0
\end{equation}
where the second and the third terms on the left take into account the factor ordering ambiguities in \eqref{final hamiltonian constraint}. We assume that the wave functional is of the form 
\begin{equation}
\Psi[\tau(r),R(r),\Gamma(r)]=U\left(\int dr\Gamma(r)\frac{iW(\tau(r),R(r),\Gamma(r))}{2}\right).
\end{equation}
Since we would want to put the system on a lattice such that the wave functional can be written as the product of wave functions for each lattice point, that is $\Psi=\prod_{i}\psi_{i}$ where $i$ designates the lattice points, we take $U$ to be the exponential function. With this choice, and denoting the lattice spacing as $\sigma$, the assumed form of the wave functional becomes
\begin{eqnarray} \label{lattice wave function}
\Psi[\tau(r),R(r),\Gamma(r)] &=& e^{\int dr\Gamma(r)\frac{iW(\tau(r),R(r),\Gamma(r))}{2}} \nonumber \\
&=& e^{\lim_{\sigma\rightarrow0}\sum_{j}\sigma\Gamma_{j}\frac{iW_{j}}{2}(\tau(r_{j}),R(r_{j}),\Gamma(r_{j}))} \nonumber \\
&=& \lim_{\sigma\rightarrow0}\prod_{j}e^{\sigma\Gamma_{j}\frac{iW_{j}}{2}(\tau(r_{j}),R(r_{j}),\Gamma(r_{j}))} \nonumber  \\
&=& \lim_{\sigma\rightarrow0}\prod_{j}\psi_{j}(\tau(r_{j}),R(r_{j}),\Gamma(r_{j})).
\end{eqnarray}
Here we recall that $\Gamma=nF'/\tilde{G}$ which implies $F(r)$ is given by \eqref{definition of F} and therefore on the lattice we have $F(r_{i})=\lim_{\sigma\rightarrow0}\sum_{j=0}^{i}\tG\sigma\Gamma_{j}/n$. We also have $\delta(0)\rightarrow\lim_{\sigma\rightarrow0}1/\sigma$. Now the lattice version of the WDW equation is \cite{wd}
\begin{equation} \label{lattice wdw eq}
\left[\hbar^{2}\left(\frac{\partial^{2}}{\partial\tau_{j}^{2}}+\mathcal{F}_{j}\frac{\partial^{2}}{\partial R_{j}^{2}}+A(R_{j},F_{j})\frac{\partial}{\partial R_{j}}+B(R_{j},F_{j})\right)+\frac{\sigma^{2}\Gamma_{j}^{2}}{\mathcal{F}_{j}}\right]\psi_{j}=0.
\end{equation}
After substituting for $\psi_{j}$ from (\ref{lattice wave function}) the above equation gives
\begin{equation}
-\frac{\sigma^{2}\Gamma_{j}^{2}}{4}\left[\hbar^{2}\left(\frac{\partial W_{j}}{\partial\tau_{j}}\right)^{2}+\hbar^{2}\mathcal{F}_{j}\left(\frac{\partial W_{j}}{\partial\ R_{j}}\right)^{2}-\frac{4}{\mathcal{F}_{j}}\right]+\frac{i\sigma\Gamma_{j}}{2}\left[\hbar^{2}\frac{\partial^{2}W_{j}}{\partial\tau_{j}^{2}}+\hbar^{2}\mathcal{F}_{j}\frac{\partial^{2}W_{j}}{\partial R_{j}^{2}}+\hbar^{2}A\frac{\partial W_{j}}{\partial R_{j}}\right]+\hbar^{2}B(R_{j},F_{j})=0.
\end{equation}
For the above equation to be satisfied independent of the choice of $\sigma$ the following three equations must be satisfied \cite{k3}
\begin{equation} \label{sigma square eq}
\left(\hbar\frac{\partial W}{\partial\tau}\right)^{2}+\mathcal{F}\left(\hbar\frac{\partial W}{\partial\ R}\right)^{2}-\frac{4}{\mathcal{F}}=0,
\end{equation}
\begin{equation} \label{sigma eq}
\left(\frac{\partial^{2}}{\partial\tau^{2}}+\mathcal{F}\frac{\partial^{2}}{\partial R^{2}}+A\frac{\partial}{\partial R}\right)W=0,
\end{equation}
\begin{equation}
B=0,
\end{equation}
where the label $j$ has been ignored in writing the above equations, it being understood that these equations are to be satisfied at each lattice point. To solve the above equations we look for  solutions of the form $W(\tau,R)=\alpha(\tau)+\beta(R)$. With this (\ref{sigma square eq}) and (\ref{sigma eq}) respectively become
\begin{equation} \label{first eq for alpha beta} 
\hbar^{2}\left(\frac{d\alpha}{d\tau}\right)^{2}+\hbar^{2}\mathcal{F}\left(\frac{d\beta}{d R}\right)^{2}-\frac{4}{\mathcal{F}}=0,
\end{equation}
\begin{equation} \label{second eq for alpha beta}
\frac{d^{2}\alpha}{d\tau^{2}}+\mathcal{F}\frac{d^{2}\beta}{d R^{2}}+A\frac{d\beta}{dR}=0.
\end{equation}
In both the above equations the first term is dependent only on $\tau$ while the rest of the terms are independent of it (depending only on $\Gamma$ and $R$), which therefore implies that these must be constant. With this \eqref{first eq for alpha beta} implies $\hbar^{2}\left(\frac{d\alpha}{d\tau}\right)^{2}=c_{1}^{2}$ where $c_{1}$ is a constant. The solution of this is $\alpha=\pm(c_{1}\tau+c_{0})/\hbar$.
Using this solution for $\alpha$, (\ref{first eq for alpha beta}) becomes
\begin{equation}
c_{1}^{2}+\hbar^{2}\mathcal{F}\left(\frac{d\beta}{d R}\right)^{2}-\frac{4}{\mathcal{F}}=0.
\end{equation}
From this we get
\begin{equation}
\frac{d\beta}{d R}=\pm\frac{\sqrt{4-c_{1}^{2}\mathcal{F}}}{\hbar\mathcal{F}}.
\end{equation}
This can be differentiated with respect to $R$ to obtain
\begin{equation}
\frac{d^{2}\beta}{dR^{2}}=\pm\frac{1}{\hbar\mathcal{F}^{2}}\left[-\frac{c_{1}^{2}\mathcal{F}}{2\sqrt{4-c_{1}^{2}\mathcal{F}}}-\sqrt{4-c_{1}^{2}\mathcal{F}}\right]\frac{d\mathcal{F}}{dR}.
\end{equation}  
 Also from the solution for $\alpha$ we see that the first term of (\ref{second eq for alpha beta}) is zero which therefore implies
\begin{equation}
\mathcal{F}\frac{d^{2}\beta}{d R^{2}}+A\frac{d\beta}{dR}=0
\end{equation}
Substituting for $d\beta/dR$ and $d^{2}\beta/dR^{2}$ in the above equation we get
\begin{equation}
2A(4-c_{1}^{2}\mathcal{F})=[c_{1}^{2}\mathcal{F}+2(4-c_{1}^{2}\mathcal{F})]\frac{d\mathcal{F}}{dR}.
\end{equation}
This equation when solved for $A$ gives
\begin{equation}
A=\frac{\left(\frac{(n-1)F}{R^{n}}+\frac{4\Lambda R}{n(n+1)}\right)\left(8-c_{1}^{2}+\frac{c_{1}^{2}F}{R^{n-1}}-\frac{2c_{1}^{2}\Lambda R^{2}}{n(n+1)}\right)}{\left(8-2c_{1}^{2}+\frac{2c_{1}^{2}F}{R^{n-1}}-\frac{4c_{1}^{2}\Lambda R^{2}}{n(n+1)}\right)}.
\end{equation}
Thus for this $A$ we get a separable solution for $W$
\begin{equation} \label{solution for w}
W=\frac{1}{\hbar}\left(c_{0}\pm c_{1}\tau\pm\int dR\frac{\sqrt{4-c_{1}^{2}\mathcal{F}}}{\mathcal{F}}\right).
\end{equation}
With this form for $W$ the wave function becomes
\begin{equation} \label{solution for psi}
\Psi[\tau(r),R(r),\Gamma(r)]=e^{\int dr\frac{i}{2\hbar}\Gamma(r)\left(c_{0}\pm c_{1}\tau\pm\int dR\frac{\sqrt{4-c_{1}^{2}\mathcal{F}}}{\mathcal{F}}\right)}.
\end{equation}
Before proceeding further we would like to express the dust proper time $\tau$ in terms of the Killing time $T$. From \eqref{tau prime in presence of E} we recall that (putting $E=0$, as is the case for marginally bound model) 
\begin{equation} \label{tau prime}
\tau'=\frac{R'}{\mathcal{F}}\sqrt{1-\mathcal{F}}-\frac{\tilde{G}_{d}LP_{L}}{2n\mathcal{F}R^{n-1}}.
\end{equation}
However, we also have $P_{F}=LP_{L}/2R^{n-1}\mathcal{F}=-nP'_{\Gamma}/\tilde{G}_{d}$, which implies that $\tau$ can be written as
\begin{equation}
\tau=P_{\Gamma}\mp\int dR\frac{\sqrt{1-\mathcal{F}}}{\mathcal{F}}.
\end{equation}
Now in \cite{kuchar} it was shown that for the Schwarzschild geometry $P_{\Gamma}$ equals the Killing time $T$ (there it is $2P_{\Gamma}=T$; however, recall that we have absorbed $n$ in the definition of $P_{\Gamma}$). Therefore if we have small dust perturbations not affecting the spacetime geometry, this relation may still be used and we get 
\begin{equation}
\tau=T\mp\int dR\frac{\sqrt{1-\mathcal{F}}}{\mathcal{F}}.
\end{equation} 
We note that here the minus sign is for expanding dust cloud and the plus sign is for the collapsing dust cloud. Now we can also fix the value of the constant $c_{1}$ appearing in the expression for the wave function. For this we note that the Hamiltonian constraint \eqref{final hamiltonian constraint} can be solved for $\bar{P}_{R}$ giving
\begin{equation}
\bar{P}_{R}=\pm\frac{P_{\tau}}{\mathcal{F}}\sqrt{\frac{\Gamma^{2}}{P_{\tau}^{2}}-\mathcal{F}}.
\end{equation}
This when substitued in the diffeomorphism constraint \eqref{final diffeo constraint} gives
\begin{equation}
\tau'\pm\frac{R'}{\mathcal{F}}\sqrt{\frac{\Gamma^{2}}{P_{\tau}^{2}}-\mathcal{F}}-\frac{\Gamma P'_{\Gamma}}{P_{\tau}}\approx0.
\end{equation}
Comparing this with \eqref{tau prime} after expressing it in terms of $P_{F}$ we have that $P_{\tau}=\Gamma$. This implies that $\hat{P}_{\tau}\Psi=\hat{\Gamma}\Psi$. Evaluating the two sides we get $c_{1}=2$. 

 Using this in (\ref{solution for psi}) the wave function becomes
\begin{equation} \label{wave function in terms of killing time}
\Psi[\tau(r),R(r),\Gamma(r)]=e^{\int dr\frac{i}{\hbar}\Gamma(r)\left[\frac{c_{0}}{2}\pm \left(T\mp\int dR\frac{\sqrt{1-\mathcal{F}}}{\mathcal{F}}\right)\pm\int dR\frac{\sqrt{1-\mathcal{F}}}{\mathcal{F}}\right]}.
\end{equation}
We are interested in calculating the Bogolubov coefficient in the near horizon limit outside the horizon for which $\mathcal{F}>0$. The Bogolubov coefficient is given by
\begin{equation} \label{bogolubov coefficient}
\beta_{\omega\omega'}=\frac{\sigma\omega}{\tG_{d}\hbar}\int_{R_{eh}}^{\infty}dR\sqrt{g_{RR}}\Psi^{*}_{og}(\omega)\Psi_{ig}(\omega')
\end{equation}
Here $\sigma\omega/\tG_{d}\hbar$ replaces the $2\omega$ in the standard definition of the Bogoliubov coefficient (see below for further explanation). Subscripts og/ig stand for outgoing/ingoing modes and $R_{eh}$ is the area radius at the horizon. Now $\sqrt{g_{RR}}$ is the metric coefficient on the space of metrics. From the form of the Hamiltonian constraint this metric is seen to be diagonal in the $(\tau,R)$ coordinates. In the $(T,R)$ coordinates its form can be determined by using
\begin{equation}
g_{RR}(T,R)=g_{\tau\tau}\left(\frac{d\tau}{dR}\right)^{2}+g_{RR}\left(\frac{dR}{dR}\right)^{2}
\end{equation}
which yields $\sqrt{g_{RR}}=1/\mathcal{F}$.
For an expanding dust cloud the ingoing wave function is given by
\begin{equation} \label{ingoing mode}
\Psi^{+}_{ig}[\tau,R,\Gamma]=e^{\frac{i}{\hbar}\int dr\Gamma(r)(\frac{c_{0}}{2}+T)}.
\end{equation}
Similarly the outgoing mode is given by
\begin{equation} \label{outgoing mode}
\Psi^{-}_{og}[\tau,R,\Gamma]=e^{\frac{i}{\hbar}\int dr\Gamma(r)(\frac{c_{0}}{2}-T+2\int dR\frac{\sqrt{1-\mathcal{F}}}{\mathcal{F}})}.
\end{equation}
On the lattice the above modes become
\begin{eqnarray} \label{ingoing and outgoing modes on lattice}
\psi^{+}_{ig} &=& \lim_{\sigma\rightarrow0}\prod_{i}e^{i\frac{\sigma}{\tG_{d}\hbar}(\frac{b_{i}}{2}+T_{i})\omega_{i}} \\
\psi^{-}_{og} &=& \lim_{\sigma\rightarrow0}\prod_{i}e^{i\frac{\sigma}{\tG_{d}\hbar}(\frac{b_{i}}{2}-T_{i}+2\int dR\frac{\sqrt{1-\mathcal{F}}}{\mathcal{F}})\omega_{i}}
\end{eqnarray}
Here we have replaced $c_{0}$ by $b$ and  we have defined $n\omega=F'$, which implies $\Gamma=\omega/\tG_{d}$. With everything set we now evaluate (\ref{bogolubov coefficient}) by substituting for the modes (in the following we will drop the subscript $i$ always remembering that we are working on a lattice).
\begin{equation} \label{bogolubov coefficient in terms of f}
\beta_{\omega\omega'}=\frac{\sigma\omega}{\tG_{d}\hbar}\exp\left(\frac{i\sigma}{\tG_{d}\hbar}[\frac{b}{2}(\omega'-\omega)+T(\omega'+\omega)]\right)\int_{R_{eh}}^{\infty}dR\frac{1}{\mathcal{F}}\exp\left(-\frac{i2\sigma\omega}{\tG_{d}\hbar}\int dR\frac{\sqrt{1-\mathcal{F}}}{\mathcal{F}}\right).
\end{equation} 
As a first step we calculate the Bogoliubov coefficient in the absence of the cosmological constant in which case $\mathcal{F}=1-F/R^{n-1}$. We do a near horizon calculation noting that at the horizon $1-F/R^{n-1}=0$. To simplify the integration we define
\begin{equation}
s=\sqrt{\frac{R^{n-1}}{F}}-1.
\end{equation}
With this the integral term in the above equation becomes
\begin{equation}
\frac{2F^{\frac{1}{n-1}}}{n-1}\int_{0}^{\infty}ds\frac{(1+s)^{\frac{n+1}{n-1}}}{s^{2}+2s}\exp\left(-i\frac{4\sigma\omega F^{\frac{1}{n-1}}}{\tG_{d}\hbar(n-1)}\int^{s}d\bar{s}\frac{(1+\bar{s})^{\frac{2}{n-1}}}{\bar{s}^{2}+2\bar{s}}\right).
\end{equation}
To regularize the integral at infinity we multiply it with $e^{-ps}$, $p>0$ and take the limit $p\rightarrow0$ after doing the integration. Thus there is a suppression for large values of $s$ and we can assume that $s$ is small. Thus retaining only the first order terms in $s$ the integral becomes
\begin{equation}
\lim_{p\rightarrow0}\frac{F^{\frac{1}{n-1}}}{n-1}\int_{0}^{\infty}ds\left(s^{-1-\frac{i2\sigma\omega F^{1/n-1}}{\tG_{d}\hbar(n-1)}}e^{-\frac{i4\sigma\omega F^{1/n-1}s}{\tG_{d}\hbar(n-1)^{2}}}e^{-ps}\right).
\end{equation}
The above integral evaluates to
\begin{equation}
\frac{F^{\frac{1}{n-1}}}{n-1}\left[\frac{4\sigma\omega F^{\frac{1}{n-1}}s}{\hbar(n-1)^{2}}\right]^{\frac{i2\sigma\omega F^{1/n-1}}{\tG_{d}\hbar(n-1)}}e^{-\frac{\pi \sigma\omega F^{1/n-1}}{\tG_{d}\hbar(n-1)}}\Gamma\left(-\frac{i2\sigma\omega F^{1/n-1}}{\tG_{d}\hbar(n-1)}\right).
\end{equation} 
With this the Bogoliubov coefficient turns out to be
\begin{eqnarray}
\beta_{\omega\omega'} &=& \frac{\sigma\omega}{\tG_{d}\hbar}\frac{F^{\frac{1}{n-1}}}{n-1}\exp\left(\frac{i\sigma}{\tG_{d}\hbar}[\frac{b}{2}(\omega'-\omega)+T(\omega'+\omega)]\right)\left[\frac{4\sigma\omega F^{\frac{1}{n-1}}s}{\tG_{d}\hbar(n-1)^{2}}\right]^{\frac{i2\sigma\omega F^{1/n-1}}{\tG_{d}\hbar(n-1)}} \nonumber \\
&& e^{-\frac{\pi \sigma\omega F^{1/n-1}}{\tG_{d}\hbar(n-1)}}\Gamma\left(-\frac{i2\sigma\omega F^{1/n-1}}{\tG_{d}\hbar(n-1)}\right).
\end{eqnarray} 
Using $|\Gamma(iy)|^{2}=\pi/y\sinh(\pi y)$, (y real), and noting that particle creation rate corresponds to the absolute square of the Bogoliubov coefficient we find that 
\begin{equation}
|\beta_{\omega\omega'}|^{2}=\left(\frac{\pi\sigma\omega F^{1/n-1}}{\tG_{d}\hbar}\right)\frac{1}{e^{\frac{4\pi\sigma\omega F^{1/n-1}}{\tG_{d}\hbar(n-1)}}-1}.
\end{equation}
By defining $\sigma\omega\equiv\tG_{d}\Delta E$, $\Delta E$ being the energy of a shell, this can be interpreted as an eternal Schwarzschild black hole in equilibrium with a thermal bath at the Hawking temperature
\begin{equation}
k_{B}T_{H}=\frac{(n-1)\hbar}{4\pi F^{1/n-1}}.
\end{equation}
For instance in $3+1$ dimensions where $n=2$ we find
\begin{equation}
k_{B}T_{H}=\frac{\hbar}{8\pi GM}.
\end{equation}

We now calculate the Bogoliubov coefficient in the presence of the negative cosmological constant. We again work in the near horizon approximation. We note that the horizon in \eqref{bogolubov coefficient in terms of f} is given by the root of $\mathcal{F}$
\begin{equation}
\mathcal{F}=1-\frac{F}{R^{n-1}}+\frac{2\Lambda R^{2}}{n(n+1)}=0.
\end{equation}
Since this equation cannot be solved generally for arbitrary $n$, we write the expression for $\mathcal{F}$ in terms of the surface gravity. The surface gravity is given by (for an exact expression of surface gravity see the following section)
\begin{equation}
\kappa=-\frac{1}{2}\frac{dg_{00}}{dR}\vert_{R=R_{eh}}
\end{equation}
However, $g_{00}=\mathcal{F}$ which, as seen above, equals zero at the horizon and therefore we can approximate $\mathcal{F}$ near the horizon by
\begin{equation}
\mathcal{F}=-2\kappa(R-R_{eh}).
\end{equation}
Defining $R-R_{eh}:=s$ we have that $\mathcal{F}=-2\kappa s$ and with this the expression for Bogoliubov coefficient becomes
\begin{equation}
\beta_{\omega\omega'}=\frac{\sigma\omega}{\tG_{d}\hbar}\exp\left(\frac{i\sigma}{\tG_{d}\hbar}[\frac{b}{2}(\omega'-\omega)+T(\omega'+\omega)]\right)\int_{0}^{\infty}ds\left(\frac{-1}{2\kappa s}\right)\exp\left(-\frac{i2\sigma\omega}{\tG_{d}\hbar}\int^{s} d\bar{s}\frac{\sqrt{1+2\kappa\bar{s}}}{-2\kappa\bar{s}}\right).
\end{equation}
To regularize the integral we multiply it with $s^{q}e^{-ps}$, $(p,q)>0$ (and take the limit $(p,q)\rightarrow0$ after doing the integral) which  would imply that we will get contributions only for those values of $s$ which are near zero. For this reason we retain only the first order terms in the integral in the exponential. With this the above integral becomes
\begin{equation}
\beta_{\omega\omega'}=\lim_{p,q\rightarrow0}\left[-\frac{\sigma\omega}{2\kappa \tG_{d}\hbar}\exp\left(\frac{i\sigma}{\tG_{d}\hbar}[\frac{b}{2}(\omega'-\omega)+T(\omega'+\omega)]\right)\int_{0}^{\infty}ds\left(s^{-1+\frac{i\sigma\omega}{\tG_{d}\hbar\kappa}}e^{\frac{i\sigma\omega}{\tG_{d}\hbar\kappa}s}s^{q}e^{-ps}\right)\right].
\end{equation}
This finally evaluates to 
\begin{equation}
\beta_{\omega\omega'}=-\frac{\sigma\omega}{2\kappa \tG_{d}\hbar}\exp\left(\frac{i\sigma}{\tG_{d}\hbar}[\frac{b}{2}(\omega'-\omega)+T(\omega'+\omega)]\right)\left(-\frac{i\sigma\omega}{\tG_{d}\hbar\kappa}\right)^{-\frac{i\sigma\omega}{\tG_{d}\hbar\kappa}}\Gamma\left(-\frac{i\sigma\omega}{\tG_{d}\hbar\kappa}\right).
\end{equation}
As before we are interested in the particle creation rate for which we take the absolute square of the above equation which gives
\begin{equation}
|\beta_{\omega\omega'}|^{2}=\left(\frac{\pi\sigma\omega}{2\tG_{d}\hbar\kappa}\right)^{2}\left(\frac{1}{e^{\frac{2\pi \sigma\omega}{\tG_{d}\hbar\kappa}}-1}\right).
\end{equation} 
As before, using the correspondence $\sigma\omega\equiv \tG_{d}\Delta E$, we have a black hole in equilibrium with a thermal bath at Hawking temperature 
\begin{equation}
k_{B}T_{H}=\frac{\hbar\kappa}{2\pi}.
\end{equation}
Since it is difficult to solve for the horizon radius for an arbitrary $n$ we cannot give a general formula for the surface gravity and hence for the Hawking temperature which will show its dependence on the number of dimensions, the mass of the black hole and the value of the cosmological constant. However, in the next section we see how these three quantities control the behaviour of the Hawking temperature and the specific heat.

\section{Surface Gravity}

As follows from (\ref{exterior metric for negative lambda}) the event horizon is defined by the condition
\begin{equation} \label{condition for eh negative lambda}
1-\frac{F_{s}}{x^{n-1}}+\frac{2\Lambda x^{2}}{n(n+1)}=0.
\end{equation}
This implies that the radius of the event horizon is given as the solution to the following equation
\begin{equation} \label{radius of eh negative lambda}
2\Lambda x^{n+1}+n(n+1)x^{n-1}-n(n+1)F_{s}=0.
\end{equation}
In particular, for $n=2$ we find that the event horizon is given by
\begin{equation} \label{radius of eh in 4d negative lambda}
x_{eh}=-\frac{1}{(3GM\Lambda^{2}+\sqrt{\Lambda^{3}+9G^{2}M^{2}\Lambda^{4}})^{1/3}}+
\frac{(3GM\Lambda^{2}+\sqrt{\Lambda^{3}+9G^{2}M^{2}\Lambda^{4}})^{1/3}}{\Lambda}.
\end{equation}
Area of the event horizon is nothing but the volume of an $n$-sphere of radius $x_{eh}$ which is given by
\begin{equation} \label{area of eh}
A_{eh}=\frac{2\pi^{\frac{n+1}{2}}}{\Gamma(\frac{n+1}{2})}x_{eh}^{n}
\end{equation}
where $2\pi^{\frac{n+1}{2}}/\Gamma(\frac{n+1}{2})$ is the volume of a unit $n$-sphere. The surface gravity is given by 
\begin{equation} \label{formula for surface gravity}
\kappa=-\frac{1}{2}\frac{dg_{00}}{dx}|_{x=x_{eh}}
\end{equation}
and works out to
\begin{equation}
\kappa=\frac{1}{2}\left[\frac{(n-1)F_{s}}{x_{eh}^{n}}+\frac{4\Lambda x_{eh}}{n(n+1)}\right].
\end{equation}
Using $F_{s}=(1+2\Lambda x_{eh}^{2}/n(n+1))x_{eh}^{n-1}$ we get
\begin{equation} \label{surface gravity negative lambda}
\kappa=\frac{1}{2}\left[\frac{n-1}{x_{eh}}+\frac{2\Lambda x_{eh}}{n}\right].
\end{equation}
Noting that the Hawking temperature of the quantized black hole is determined by surface gravity, we see that this expression for surface gravity has remarkable implications for black hole thermodynamics. The Hawking temperature is controlled by three parameters: mass of the black hole, number of spatial dimensions and the cosmological constant. The presence of $\Lambda$ could make the specific heat of the black hole positive (as it does for the BTZ black hole). To this effect we calculate the derivative $d\kappa/dM$ for fixed $n$ and $\Lambda$. From (\ref{surface gravity negative lambda}) we get that
\begin{equation}
\label{spheat}
2\frac{d\kappa}{dM} = \frac{dx}{dM}\left[-\frac{n-1}{x^2} + \frac{2\Lambda}{n}\right]
\end{equation}
where to simplify notation we write $x$ instead of $x_{eh}$. As is expected, and also evident from
(\ref{radius of eh negative lambda}), $x$ increases monotonically with $M$. Thus the condition for positivity of the specific heat is
\begin{equation}
\label{psh}
2\Lambda x^2  > n(n-1)
\end{equation}
which is a constraint on the dimensionless quantity constructed from the cosmological constant and the radius of the event horizon. The above relation was also obtained in \cite{sdas}. After expressing $x$ in terms of $M$, $\Lambda$ and $n$, this becomes a condition on the mutual relation between the three free parameters. A few things can be read off immediately: (i) if $\Lambda$ is zero, specific heat is necessarily negative, (ii) for the 4-d AdS case, i.e. $n=2$, specific heat is positive if $\Lambda x^2 >1$. Using the expression for $x$ from (\ref{radius of eh in 4d negative lambda}) this appears to translate into a complicated relation between $M$ and $\Lambda$, (iii) given any $n>1$, and some value of $\Lambda$, one can always choose an $M$, and hence an $x$, sufficiently large, so that the specific heat becomes positive. 

\section{Conclusions}

We have used the canonical formalism to study the quantization of a spherically symmetric dust cloud in the presence of a negative cosmological constant in arbitrary number of spatial dimensions. It is remarkable that the canonical form is same in all dimensions, and is also not affected by the inclusion of the cosmological constant.  We obtained the Wheeler-DeWitt equation which was solved to obtain the exact quantum states using lattice regularization. In the limit of small dust perturbations over the AdS-Schwarzschild geometry these states correspond to Hawking radiation. This provides justification for the choice of the inner product on the space of metrics. We also saw that there is an interesting interplay between the number of spatial dimensions, the cosmological constant and the mass of the black hole, which can render the specific heat of the black hole positive. 

It would also be interesting to see the role of this interplay in determining the entropy of an AdS-Schwarzschild black hole in arbitrary number of dimensions. This aspect will be discussed in the next paper where we calculate the entropy by counting the microscopic degrees of freedom using the quantized states obtained here. The change of the sign of the specific heat is likely to have important consequences for the statistical description of the thermodynamics. 

\section*{Acknowledgements}

R.T. thanks Sashideep Gutti for useful discussions and S. Shankarnarayanan for suggesting one of the references.


\begin{thebibliography}{99}
\bibitem{hawking} S. W. Hawking, Comm. Math. Phys. 43, 199 (1975).
\bibitem{bekenstein1} Jacob D. Bekenstein, Phys. Rev. D7 (1973) 2333.
\bibitem{bekenstein2} Jacob D. Bekenstein, Phys. Rev. D9 (1974) 3292.
\bibitem{bardeen} J. M. Bardeen, B. Carter and S. W. Hawking, Comm. Math. Phys. 31, 161 (1973).
\bibitem{wd} C. Vaz, L. Witten and T. P. Singh, Phys.Rev. D63 (2001) 104020.
\bibitem{vawi1} C. Vaz and L. Witten, Phys. Rev. D60 (1999) 024009.
\bibitem{vawi2}C. Vaz and L. Witten, Phys. Rev. D63 (2001) 024008.
\bibitem{k1} C. Vaz, C. Kiefer, T. P. Singh and L. Witten,
 Phys. Rev. D67 (2003) 024014.
\bibitem{k2} C. Vaz, L. Witten and T. P. Singh, Phys. Rev. D69 (2004) 104029. 
\bibitem{k3} C. Kiefer, J. Mueller-Hill and C. Vaz, Phys. Rev. D73, 044025 (2006).
\bibitem{k4} C. Kiefer, J. Mueller-Hill, T. P. Singh and C. Vaz, Phys. Rev. D75, 124010 (2007).
\bibitem{kuchar} Karel V. Kucha\v r, Phys. Rev. D50, 3961 (1994).
\bibitem{sashi1} Cenalo Vaz, Sashideep Gutti, Claus Kiefer and T. P. Singh, Phys. Rev. D76, 124021 (2007).
\bibitem{sashi2} Cenalo Vaz, Sashideep Gutti, Claus Kiefer, T. P. Singh and L. C. R. Wijewardhana, Phys. Rev. D77, 064021 (2008).
\bibitem{rtibs} Rakesh Tibrewala, Sashideep Gutti, T. P. Singh and Cenalo Vaz, Phys. Rev. D77, 064012 (2008).
\bibitem{torre} K.V. Kucha\v r and C. G. Torre, Phys. Rev. D43, 419 (1991).
\bibitem{brown} J. D. Brown and K. V. Kucha\v r, Phys. Rev. D51, 5600 (1995).
\bibitem{sdas} Saurya Das, Parthasarathi Majumdar and Rajat K. Bhaduri, Class. Quant. Grav. 19, 2355 (2002).
\end{thebibliography}
\end{document}